\documentclass[aps,prd,twocolumn,showpacs,showkeys,superscriptaddress]{revtex4-2}
\usepackage{latexsym}
\usepackage{amssymb}
\usepackage{amsmath}
\usepackage{amscd}
\usepackage{amsthm}
\usepackage{graphicx}
\usepackage{textcomp}
\usepackage{colortbl}
\usepackage{hyperref}
\usepackage[font={footnotesize,it}]{caption}
\usepackage{multirow}
\usepackage[T1]{fontenc}
\usepackage{ae,aecompl}

\begin{document}

\title{Reexamining RHDE Models in FRW Universe with two IR cutoff with redshift parametrization}
\author{Anil Kumar Yadav}
\email{abanilyadav@yahoo.co.in}
\affiliation{Department of Physics, United College of Engineering and Research,Greater Noida - 201310, India}
 
\begin{abstract}
In this paper, we have investigated that the gravitational field equations are not compatible with conservation equation in Dixit et al. \textcolor{blue}{[Dixit et al. Euro. Phys. J. Plus \textbf{135}, 831 (2020)]}. Therefore, the expression for equation of state parameter along with the dynamics of $\omega_{T}$ - $\omega_{T}^{\prime}$ plane  does not reflect the actual behaviors of RHDE models in $f(R,T)$ gravity and thus the method and technique given in Dixit et al. \textcolor{blue}{[Dixit et al. Euro. Phys. J. Plus \textbf{135}, 831 (2020)]} represents a fractured way for analyzing RHDE models in $f(R,T)$. We also investigate that the derived Universe is in decelerating phase of expansion for $0 \leq \beta \leq 1.5$ which is contrary to the result obtained in above targeted paper.     
\end{abstract}



\maketitle

\section{Introduction}
\label{sec:intro}
In 2004, Li \cite{Li/2004} has proposed holographic dark energy principle to search the dark energy scenario and the authors of Refs. \cite{Nojiri/2006,Li/2009,Gong/2005} have investigated that this concept of holographic dark energy can be utilize in quantum gravity. In Wang and Wang \cite{Wang/2017}, a holographic dark energy model inspired by Bekenstein - Hawking entropy has been investigated. We also note that, by using the concept of horizon entropy of a black hole which is also known as Tsallis entropy \cite{Tsallis/2013}, Tavayef et al. \cite{Tavayef/2018} have investigated Tsallis holographic dark energy (THDE) model in general relativity. However, the Hubble horizon as an IR cutoff in modified theories of gravity is not a suitable candidate to explain the late time acceleration in the Universe \cite{Xu/2009,Yadav/2020}. Some important applications of Tsallis entropy are given in Ref. \cite{Nunes/2016}.\\

Recently, Moradpour et al. \cite{Moradpour/2018} have proposed a new holographic model for dark energy model inspired by the concept of R$\grave{e}$nyi entropy \cite{Renyi/1961} in the framework of general relativity. This new model of dark energy is known as R$\grave{e}$nyi holographic dark energy (RHDE) model. In the recent past, Tsallis and R$\grave{e}$nyi entropies \cite{Masi/2005,Touchette/2002,Tsallis/2011,Renyi/1970,Tsallis/1988} have been used in order to study some gravitational events in the framework of general relativity \cite{Komatsu/2017,Moradpour/2017,Moradpour/2018plb,Moradpour/2016,Abreu/2013,Abreu/2013a}. Also, we note that a general approach to THDE and even its generalization is done in Refs. \cite{Nojiri/2019,Nojiri/2020}.\\ 

In the recent time, the modified theories of gravity have been constantly used to solve dark energy problems or issues associated with standard $\Lambda$CDM model \cite{Padmanabhan/2008,Yoo/2012,Wang/2006}. Some alternative theories of gravity have pretty agreement with astrophysical observations \cite{Demianski/2006,Lubini/2011,Deng/2015}. The 
$f(R,T)$ theory of gravity \cite{Harko/2011} presents in its field equations extra contributions from both geometry, through a general dependence on $R$, and matter, through a general dependence on $T$, the trace of the energy-momentum tensor. Moreover, the T - dependence of the geometrical action in $f(R,T)$ gravity may be due to the existence of some imperfect fluids and intrinsically may have some quantum effects like particle production \cite{Harko/2014}. Some important applications of $f(R,T)$ theory of gravity in various physical contexts are given in Refs. \cite{Singh/2014,Houndjo/2014,Sharif/2013,Moraes/2016,Moraes/2015,Correa/2016,Yadav/2019,Yadav/2018,Yadav/2014,Prasad/2020}. \\

The article is organized as follows: The theoretical model and its mathematical formalism is given in section \ref{sec:2}. In Section \ref{sec:3} the violation/validation of conservation law for RHDE models in $f(R,T)$ theory of gravity are discussed. Finally, in section \ref{sec:4}, we summarized our findings.
\section{Theoretical model and Basic equations}
\label{sec:2}
The modified Einstein - Hilbert action for the $f(R, T )$ theory of gravity
is read as \cite{Harko/2011}
\begin{equation}
\label{A-1}
S = \int\sqrt{-g}\left[\frac{1}{16\pi G}f(R,T)+\mathcal{L}_{m}\right]dx^{4}.
\end{equation}
where $\mathcal{L}_{m}$ is the matter Lagrangian density of matter and $f(R,T)$ is an arbitrary function of $R$ and $T$. The other symbols have their usual meanings.\\
The energy momentum tensor $ T_{i j} $ of matter is written as
\begin{equation}\label{intro2}
T_{i j}=-\frac{2}{\sqrt{-g}}\frac{\delta\left(\sqrt{-g}\mathcal{L}_{m}\right)}{\delta g^{i j}}.
\end{equation}
and its trace $T = g^{i j}  T_{ i j} $.\\

In Dixit et al. \cite{Dixit/2020}, the authors have considered the stress-energy tensor as
\begin{equation}
\label{A-2}
T_{ij} = -pg_{ij}+(\rho + p)u_{i}u_{j}.
\end{equation}
With the choice of $\mathcal{L}_{m} = - p$, with $p$ being the pressure, and assuming units such that G = 1, the gravitational field equation for $f(R,T) = R + 2f(T)$ is read as
\begin{equation}
\label{A-3}
2R_{ij}-Rg_{ij} = 16\pi T_{ij}+4\dot{f}(T)T_{ij}+2\left[2p\dot{f}(T)+f(T)\right]g_{ij},
\end{equation}
where $\dot{f}(T) = \frac{\partial f}{\partial t}$.\\

For dust filled Universe, $p = 0$, Eq. (\ref{A-3}) is reduced to
\begin{equation}
\label{A-4}
2R_{ij}-Rg_{ij} = 16\pi T_{ij}+4\dot{f}(T)T_{ij}+2f(T)g_{ij}.
\end{equation}
Note that Eq. (\ref{A-4}) of this manuscript is exactly same as Eq. (9) of Dixit et al \cite{Dixit/2020}. Further, in Ref. \cite{Dixit/2020}, the authors have selected $f(T) = \xi T$ with $\xi$ as a constant.\\

The spatially flat FRW Universe with a time dependent scale factor $a(t)$ is represented by following space-time
\begin{equation}
\label{A-5}
ds^{2} = dt^{2} - a^{2}(t)\left(dx^{2}+dy^{2}+dz^{2}\right).
\end{equation} 
Therefore, the general field equations for $f(R,T) = R + 2\xi T$ and the metric (\ref{A-5}) are obtained as
\begin{equation}
\label{A-6}
3\frac{\dot{a}^{2}}{a^{2}} = (8\pi +3\xi)\rho_{T},
\end{equation}
\begin{equation}
\label{A-7}
2\frac{\ddot{a}}{a}+\frac{\dot{a}^{2}}{a^{2}} = \xi \rho_{T}.
\end{equation}

We note that Eqs. (\ref{A-6}) and (\ref{A-7}) are exactly same as Eqs. (11) and (12) of Dixit et al. \cite{Dixit/2020}. With choice of $\Lambda_{eff} \propto H^{2}$, where $H = \frac{\dot{a}}{a}$ is a Hubble function, the general solution the field equations is obtained in Ref. \cite{Dixit/2020} as
\begin{equation}
\label{A-8}
H(t) = \frac{2(8\pi +3\xi)}{3t(8\pi + 2\xi)} = \frac{2\beta}{3t}.
\end{equation}
where $\beta = \frac{8\pi+3\xi}{8\pi+2\xi}$. Eq. (\ref{A-8}) of this article and Eq. (14) of Dixit et al. \cite{Dixit/2020} are same. It is worthwhile to note the solution (\ref{A-8}) is not new. It has been already given in Harko et al. \cite{Harko/2011}. Along with some other issues, the main issue in Dixit et al. \cite{Dixit/2020} is that the gravitational field equations are not compatible with energy conservation equation. Therefore, the expression for equation of state parameter along with the dynamics of $\omega_{T}$ - $\omega_{T}^{\prime}$ plane  does not reflect the actual behaviors of RHDE models in $f(R,T)$ gravity.   
\section{Non-existence of energy conservation law for RHDE in $f(R,T)$ gravity}\label{sec:3}     
In Ref. \cite{Dixit/2020}, the conservation equation is
\begin{equation}
\label{E-1}
\frac{\partial \rho_{T}}{\partial t} + 3H(\rho_{T} + p_{T}) = 0.
\end{equation}
where $\rho_{T}$ is the R$\grave{e}$nyi holographic energy density. $p_{T}$ is not defined in Ref. \cite{Dixit/2020} but we understand that it is pressure of considered fluid in connection to define equation of state parameter $\omega_{T}$. It is interesting to note that for obtaining the field equations (\ref{A-6}) and (\ref{A-7}) in $f(R,T)$ gravity [which are Eqs. (11) and (12) in Dixit et al. \cite{Dixit/2020}], the authors of Ref. \cite{Dixit/2020} have assumed $p = 0$ but in conservation equation they conveniently admit pressure without concrete physical reason behind it. However, it is well known that the conservation equation does not hold in $f(R,T)$ theory of gravity. To make it clear firstly we discuss the conservation equation in the framework of general relativity.\\

The Einstein field equations with cosmological constant $(\Lambda)$ is read as
\begin{equation}
\label{E-2}
R_{ij} - \frac{1}{2}g_{ij}R + \Lambda g_{ij} = 8\pi T_{ij}.
\end{equation}
The field equations for metric (\ref{A-5}) are written as
\begin{equation}
\label{E-3}
3\frac{\dot{a}^{2}}{a^{2}} = 8\pi\rho + \Lambda = \rho_{eff},
\end{equation}
\begin{equation}
\label{E-4}
2\frac{\ddot{a}}{a}+\frac{\dot{a}^{2}}{a^{2}} = -8\pi p + \Lambda = -p_{eff}.
\end{equation}
where $\rho_{eff} = 8\pi \rho + \Lambda$ and $p_{eff} = 8\pi p - \Lambda$. Therefore, the effective equation of state parameter is read as
\begin{equation}
\label{E-5}
\omega_{eff} = \frac{p_{eff}}{\rho_{eff}} = \frac{8\pi p - \Lambda}{8\pi \rho + \Lambda}.
\end{equation}

It is worthwhile to note that in absence of matter, $\rho = 0$ and $p = 0$, the effective equation of state parameter is equal to $\omega_{eff} = -1$ \cite{Yadav/2020prd}. This represents the standard $\Lambda$CDM model of the Universe.\\
Differentiating Eq. (\ref{E-3}) with respect to t and combining the resulting equation with Eq. (\ref{E-4}), we obtain the energy conservation equation as following 
\begin{equation}
\label{E-6}
\dot{\rho_{eff}}+3\left(\rho_{eff}+p_{eff}\right)H = 0.
\end{equation}
which implies $d(\rho_{eff}V) = -p_{eff}dV$. Note that $V = a^{3}$, the volume of the Universe and the quantity $\rho_{eff}V$ represents the total energy of the Universe. As the Universe expands, the amount of dark energy in an expanding Universe increases in proportion to the volume. While in $f(R,T)$ theory of gravity, one may get a different picture because in $f(R,T) = R + 2\xi T$ gravity, $\nabla^{i}T_{ij} \neq 0$ \cite{Harko/2011}.
\begin{equation}
\label{E-7}
\nabla^{i}T_{ij} = \frac{f_{T}(R,T)}{8\pi - f_{T}(R,T)}\left[(T_{ij}+\Theta_{ij})\nabla^{i}ln f_{T}(R,T)+\nabla^{i}\Theta_{ij}\right].
\end{equation}  
where $f_{T}(R,T) = \frac{\partial f(R,T)}{\partial T}$ and $\Theta_{ij} = g^{ij}\frac{\delta T_{ij}}{\delta g^{ij}}$.
\\
In case of $f(R,T) = R +2\xi T$, Eq. (\ref{E-7}) reduces to
\begin{equation}
\label{E-8}
\nabla^{i}T_{ij} = \frac{2\xi}{8\pi - 2\xi}\left[(T_{ij}+\Theta_{ij})\nabla^{i}ln f_{T}(R,T)+\nabla^{i}\Theta_{ij}\right].
\end{equation}

It is worthwhile to note that for $\xi = 0$, we obtain $\nabla^{i}T_{ij} = 0$, but in $f(R,T) = R + 2 \xi T$, one can not choose $\xi = 0$ because $\xi = 0$ reduces $f(R,T) = R + 2\xi T$ theory to general relativity case. However for $\xi \neq 0$, the conservation equation does not hold. Some sensible researches for violation of conservation law in $f(R,T)$ theory are given in Refs. \cite{Barrientos/2014,Sahoo/2018}. In Ref. \cite{Dixit/2020}, the conservation equation (15) can neither be obtain from Eqs. (11) and (12) nor it is true for $f(R,T)$ theory of gravity. The violation of conservation law in Dixit et al. \cite{Dixit/2020} raises a serious question on the method and approaches used for constructing RHDE models in $f(R,T)$ gravity. The authors of Ref. \cite{Dixit/2020} have analyzed the cosmological features of RHDE models through conservation equation which is not compatible with the basic field equations of the model. Also, we notice that, in Ref. \cite{Varshney/2020}, similar approach has been used for analyzing the k-essence and dilation field models inspired by THDE in f(R, T) gravity which also need some moderation to establish THDE models in modified theories of gravity.\\  
\begin{figure}[h!]
\includegraphics[width=7cm,height=6cm,angle=0]{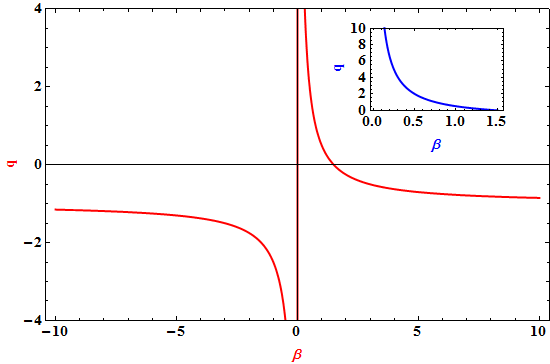}
\caption{The plot of $q$ for numerical values of $\beta$.}
\label{fig2}
\end{figure}
\section{Discussion and final remarks}\label{sec:4} 
In this paper, we have analyzed that the gravitational field equations are not compatible with energy conservation equation in the framework of $f(R,T) = R + 2\xi T$ gravity. Therefore, the method and approach given in Ref. \cite{Dixit/2020} represents a fractured way for analyzing RHDE models in $f(R,T)$. However, one can use this method to analyze some features of RHDE models in general theory of relativity. Further we investigate that the dynamics of deceleration parameter does not favor the result displayed in Dixit et al. \cite{Dixit/2020} for the particular range of $\beta$.\\

The deceleration parameter is obtained as
\begin{equation}
\label{D-1}
q = -\frac{\ddot{a}a}{\dot{a}^{2}} = -1 - \frac{\dot{H}}{H^{2}}.
\end{equation}
Eqs. (\ref{A-8}) and (\ref{D-1}) lead to
\begin{equation}
\label{D-2}
q = -1+\frac{3}{2\beta}.
\end{equation}
The plot of $q$ for numerical values of $\beta$ is shown in Fig. 1. We observe that for $0\leq \beta \leq 1.5$, the Universe in derived model is expanding in decelerating mode while in Dixit et al. \cite{Dixit/2020}, it has been classified as quintessence model of accelerating Universe for the given range of $\beta$ (see particular II of table 1 and 2 of Ref. \cite{Dixit/2020}).\\

It is worthwhile to note that neither we avoid the possible existence of RHDE cosmological models in $f(R,T)$ theory of gravity nor we decline the viability of $f(R,T)$ gravity with special type dark energy density. But the method and technique given in Dixit et al \cite{Dixit/2020} is not suitable for the dynamics of RHDE models in $f(R,T)$ gravity. As a final comment, we note that in spite of good possibility of RHDE models in $f(R,T)$ theory of gravity to provide a theoretical foundation for numerical relativity and cosmology, the experimental point is yet to be considered and still the theory needs a fair trial. \\

\end{document}